\documentclass[twocolumn]{jpsj2} 
\usepackage{graphicx, color}

\begin{document}

\title{Switching Dynamics of Bi$_{2}$Sr$_{2}$CaCu$_{2}$O$_{8+\delta}$ Intrinsic Josephson junctions: Macroscopic Quantum Tunneling and Self-Heating Effect}

\author{\textsc{Hiromi KASHIWAYA}$^{1}$\thanks{E-mail address: h.kashiwaya@aist.go.jp}, \textsc{Tetsuro MATSUMOTO}$^{1}$, \textsc{Hajime SHIBATA}$^{1}$, \textsc{Satoshi KASHIWAYA}$^{1}$, \textsc{Hiroshi EISAKI}$^{1}$, \textsc{Yoshiyuki YOSHIDA}$^{1}$, \textsc{Shiro KAWABATA}$^{2,3}$, and \textsc{Yukio Tanaka}$^{4}$}

\inst{$^{1}$Nanoelectronics Research Institute (NeRI), National Institute of Advanced Industrial Science and Technology (AIST),Tsukuba, 305-8568, Japan \\
$^{2}$Nanotechnology Research Institute, AIST, Tsukuba, 305-8568, Japan \\
$^{3}$CREST, Japan Science and Technology Corporation (JST), Kawaguchi, Saitama
332-0012, Japan\\
$^{4}$Department of Applied Physics, Nagoya University, Nagoya, 464-8603, Japan \\
}

\recdate{\today}

\abst{  
%
The switching dynamics of current-biased Bi$_{2}$Sr$_{2}$CaCu$_{2}$O$_{8+\delta}$ intrinsic Josephson junctions (IJJs) was studied to clarify the effect of $d$-wave superconductivity and the stack structure on the switching properties. 
High quality IJJs were fabricated, and then the temperature dependence of the switching probability distribution was measured for the first and second switchings.
Although the standard deviation of the distribution detected for both switchings showed similar saturation characteristics with decreasing temperature, the temperature at saturation was about 13 times higher for the second switching than for the first switching.
The properties of the first switching can be explained in terms of a conventional underdamped JJ, that is, macroscopic quantum tunneling below the crossover temperature, and thermal activation with quality factor of 70$\pm$20 above the crossover temperature.
In contrast, the relatively higher effective temperature for the second switching evaluated from the switching probability distribution suggests a dominant thermal activation process under the influence of the self-heating effect even at sufficiently low temperature.
}

\kword{intrinsic Josephson junction, macroscopic quantum phenomena, phase qubit, Bi2212}

\maketitle

\clearpage

\section{Introduction}
%
Current-biased Josephson junctions (JJs) have been accepted as an ideal system to explore the dynamics of a macroscopic quantum object under controllable environments.
The dynamics of these junctions can be equivalently represented by particle motion in a one-dimensional tilted cosine potential.
A particle is statically confined in the potential when the bias current ($I$) is sufficiently smaller than the critical current ($I_c$) and when the voltage ($V$) across the junction is zero.
When the bias current exceeds the switching current ($I_{sw}$), the particle escapes from the metastable state either by the thermal activation (TA) process or by the macroscopic quantum tunneling (MQT) process.
Various aspects of the switching properties have been examined using metal superconductor JJs, and the effect of damping due to dissipation on the switching dynamics has been clarified in numerous studies ~\cite{Voss,Devoret,Martinis1,Wallraff}.
Consequently, the quantized energy levels of current-biased JJs have recently been considered for quantum information device applications ~\cite{Han,Martinis2,Makhlin}.
On the other hand, JJs composed of high-$T_c$ superconductors (HTSCs), especially Bi$_{2}$Sr$_{2}$CaCu$_{2}$O$_{8+\delta}$ (Bi2212) intrinsic Josephson junctions (IJJs)~\cite{Kleiner}, are promising candidates for various device applications, such as oscillators and microwave sensors. 
The insulating layer of IJJs is a naturally-formed atomic layer in the crystal, and thus the insulator has high integrity compared with the artificial insulators used in metal superconductor junctions.
In addition, the relatively higher plasma frequency for IJJs compared with that for metal superconductor junctions suggests that the crossover temperature ($T^{*}$) from the classical to the quantum state should be higher as well.
These features indicate a high expectation for applications of IJJs in quantum information processing.
Actually, the MQT at relatively higher $T^{*}s$ than those of metal superconductor junctions was recently reported for HTSC JJs~\cite{Bauch,Inomata,Jin}.
However, IJJs have several inherent unique characteristics such as the $d$-wave superconductivity ~\cite{d-wave1,d-wave2}, the self-heating effect ~\cite{sh2,sh3,sh4,sh5,insitu}, the multiple stack structure, and the strong electron correlation effect.
%
How these characteristics affect the switching dynamics both in the quantum and classical regimes remains unclear.
%
To clarify this, detailed studies on the switching properties of IJJs, especially the damping effect, are needed.
%
\par
%
%
Here, the switching dynamics of IJJs was clarified by measuring the escape rate using the ramped bias current method.
One key issue in quantum device applications of HTSCs is how the peculiarities of IJJs contribute to the damping in the switching dynamics.
%
Here, the switching probability distribution $P(I)$ was measured for both the first and second switchings in a multiple-branched structure in the current-voltage characteristics.
The switching probability distribution measured for the first switching showed the conventional crossover from the TA to the MQT.
The quality (Q) factor of the first switching was then evaluated by applying statistical analysis to the TA process at a temperature ($T$) above the crossover temperature $T^{*}_1$.
The results clearly indicate underdamped features of IJJs.
The temperature dependence of the switching properties in the MQT regime observed here suggests that nodal quasiparticles (QPs) have no significant effect on the damping in the switching dynamics.
In contrast to the switching properties of the first switching, those of the second switching~\cite{secondsw} significantly differed from conventional features of an underdamped JJs.
Although $\sigma_{I}$ defined by the standard deviation of P(I) became saturated for the second switching with decreasing temperature, the detected crossover temperature $T^{*}_2$ of 6.5K was significantly higher than conventional metal superconductor cases.
We also found the reversal of the relative magnitudes of the currents corresponding to the first and second switchings,
that is, the mean switching current for the second switching ($I_{sw2}$) was higher than that for the first switching ($I_{sw1}$) when the temperature was higher than the crossover temperature ($T_{CR}$=1.1K), while $I_{sw2}$ was lower than $I_{sw1}$ when the temperature was lower than $T_{CR}$.
The self-heating effect due to the dissipative current inside the junction seriously affected the switching properties, and the self-heating assisted TA dominated the switching process in the second switching below $T^{*}_2$.
%
%
%
\section{Model}
%
In this section, we briefly introduce a model of switching dynamics of IJJs by focusing on the difference between the IJJs and metal superconductor junctions.
The JJ is described by the resistively and capacitively shunted junction (RCSJ) in which $R$ is the shunt resistance and $C$ is the shunt capacitance in the Stewart-Mccumber model ~\cite{SM}.
Because the IJJs fabricated in this study had seven JJs as described in the next section, we assumed a serially-connected RCSJ as shown in Fig. 1(a).
In the model by Koyama $et$ $al.$ ~\cite{Koyama}, each junction is coupled with neighboring junctions due to the charging effect induced by insufficient shielding of the electric field at the atomic size electrodes.
If we neglect the effect of coupling, the dynamics of a single junction can be represented by damped particle motion in a one-dimensional tilted cosine potential where the effective mass is given by $C(\Phi_{0}/2\pi$)$^2$, the damping coefficient by $1/RC$, and the magnetic flux quantum by $\Phi_{0}$ ~\cite{Remark1}.
For a constant $I$ smaller than $I_c$, the barrier height $U_0$ is given by
\begin{eqnarray}
U_0=2E_J[\sqrt{1-x^2}-x\arccos{x}],
\end{eqnarray}
where $E_{J}=\Phi_{0} I_{c}/2\pi$ and $x=I/I_{c}$.
At $T$ higher than $T^*$,
the escape rate of the particle is dominated by the TA process. 
In this case the escape rate is given by  
\begin{eqnarray}
\Gamma_{TA}=a_{t}\frac{\omega_0}{2\pi}\exp\left({-\frac{U_0}{k_bT}}\right).
\end{eqnarray}
Here, $a_t$ is the temperature- and damping-dependent thermal prefactor and
$\omega_0$($=\sqrt{2\pi I_{c}/\Phi_{0} C}(1-x^2)^{1/4}$) is the oscillation frequency at the local minima of the potential ~\cite{Kramers,Hanggi}.
The escape rate below $T^*$ ($=\hbar \omega_{0}/2\pi k_{B}$) is dominated by the MQT process, whose escape rate is given by 
\begin{eqnarray}
\Gamma_{MQT}=\sqrt{60}\omega_{0}\left(\frac{B}{2\pi}\right)^{1/2} \exp(-B),
\end{eqnarray}
where
\begin{eqnarray}
B=\frac{36U_{0}}{5\hbar \omega_{0}}(1+0.87/Q),
\end{eqnarray}
and $Q=\omega_{0} RC$ ~\cite{CL}.
The experimental escape rate $\Gamma(I)$ can be written as a function of the experimental $P(I)$ as follows,
\begin{eqnarray}
\Gamma(I)=\frac{dI}{dt}\frac{1}{\Delta I}\ln{\frac{\int^{\infty}_{I} P(I') dI' }{\int^{\infty}_{I+\Delta I}P(I') dI' }},
\end{eqnarray}
where $dI/dt$ is the sweep rate of the bias current.
The escape rate in the MQT regime is insensitive to the bath temperature when the dissipation during the tunneling process is negligible, whereas when ohmic (super-ohmic) dissipation is non-negligible, the correction to the escape rate has $T^2$ ($T^3$) dependence~\cite{Weiss}.
Several theories have been presented to estimate the effect of nodal QPs of $d$-wave superconductivity on the damping~\cite{Faminov,Amin,Joglekar,Kawabata}.
According to Kawabata $et$ $al.$ ~\cite{Kawabata}, nodal QPs induce superohmic dissipation on the MQT process reflecting the QP spectrum of the $d$-wave superconductors, although the correction is quite small.
This theory by Kawabata $et$ $al.$, however, needs to be confirmed experimentally.
Moreover it is still unclear how the switching dynamics is modified by the coupling of neighboring junctions.
Therefore, detailed study on the swhitching dynamics of IJJ is required.
\par
\section{Experimental}
%
The IJJs in this study were fabricated as follows.
First, a single crystal of Bi2212 with a critical temperature of 90K was grown by using the traveling solvent floating zone method. 
Next, the crystal was cut into 1000$\times$20$\times$5 $\mu m^3$ size samples by mechanical polishing and cleaving.
The samples were then glued onto SiO$_2$ substrates.
Four Au contact electrodes were deposited onto both ends of each sample to achieve an electrical contact to Bi2212.
A bridge structure was formed on each sample by "necking" the center region, and then an IJJ was formed by cutting the side of the sample by using a focused ion beam (FIB)~\cite{Wang}. 
The fabricated IJJ size was $1\times2\times0.01\mu m^3$, and seven JJs were stacked together to form an IJJ. 
To minimize degradation, the IJJ was immediately fixed on a sample holder installed in a dilution refrigerator.
The critical temperature ($T_c$) of the present IJJ was about 80 K.
Figure 1(b) shows a schematic of the junction configuration and a scanning ion microscope image of the IJJ. 
Based on the size of the IJJ, the estimated capacitance was between 74 (relative dielectrical constant $\epsilon _{r}$=5) and 147 fF ($\epsilon _{r}$=10).
All the electrodes were apart more than $250\mu m$ from the junction, and thus the contact resistance was less than $10\Omega$~\cite{Matsu} .
The advantage of this stacked structure over the mesa structure ~\cite{Mesa} is that the effect of the heat generated at the contacts is negligible.
Actually, no dependence of the switching probability distribution on such self-heating was detected when the duty ratio of the bias signal was decreased from 1/10 (typical value) to 1/80 at 52mK.
On the other hand, the stacked structure is rather sensitive to the heat generation because the heat-sinking channel is limited to narrow and poorly conducting arms holding the stack.
This device structure results in relatively high effective temperature at the second switching as described below.
The quality of the junction was confirmed by large hysteresis in the $I$-$V$ characteristics as shown in Fig. 2(a).
\par
%
The switching probability distribution was measured by using the time domain measurement with the ramped bias technique ~\cite{Voss,Devoret,Martinis1,Wallraff}.
Although the IJJ had four terminal configurations, negative signal lines ($V_{-}$ and $I_{-}$) were directly connected to the sample holder as shown in Fig. 2(b).
To reduce the effect of external electric noise, signal lines were carefully filtered using low-pass filters (LPF) located at room temperature, high attenuation coaxial cables, and powder microwave filters (PF) installed in the sample holder.
The cutoff frequency of the wiring was 1 MHz (or 10MHz), and an attenuation greater than 120dB was achieved at 1GHz.
A ramped bias signal was generated by using an Agilent 33220A waveform generator, and fed to the IJJ through an isolation amplifier and a bias resistor $R_I$.
The switching probability distribution was detected by using a Stanford Research time interval counter SR620 in which the time interval between the start and stop pulses from the comparators was measured.
The time interval directly corresponds to the switching voltage of the bias signal because the bias voltage was swept at a constant rate.
The distribution of the voltage was converted to that of current through the $I$-$V$ relation that was precisely measured just before the switching probability distribution measurements.
Primary result for the first switching was shown in Ref.~\citen{Matsu}.
For the measurements of second switching, we adjusted the sweep rate, the comparator threshold voltage and the preamplifier gain.
The typical sweep rate of the bias current was $\sim$ 0.011A/s and $\sim$ 0.013A/s for the first and second, respectively, and the typical sampling number to obtain one probability distribution was between 5000 and 20000.
The threshold voltage of the comparator was 2mV($=V_{th1}$) and 20mV($=V_{th2}$) for the first and the second switching, respectively (see Fig. 2(c)).
In determining the sweep rate of the current for the second switching, we approximated the $I$-$V$ characteristics of the first branch as a linear curve near the switching region (typically 18-21$\mu$A).
Estimated error due to this approximation was less than 1\%.
The time interval of about 0.2$\sim$1$\times$10$^{-3}$[sec] between the first switching and the second switching is enough to avoid the dynamical effects. 
We have checked it by changing the sweep rate of the bias voltage.
All analog circuits were isolated from the digital apparatus by optical fiber links to isolate the circuits from the digital noise sources and to avoid the ground loop.
The temperature of the sample $T_{bath}$ was monitored by using a RuO$_2$ thermometer tightly fixed to a sample holder made of oxygen-free copper.
Deviation in $T_{bath}$ from the actual sample temperature was minimized by fixing the thermometer to the sample holder just behind the sample. 
Further details of the measurement setup were reported elsewhere~\cite{HKash}.
\par
\section{Results}
%
The $I$-$V$ characteristics of the IJJs exhibited the multiple-branched structure due to successive switching of the stacked JJs in the IJJ from zero to finite voltage states as shown in Fig. 2(a).
In this study, we focused on the first and second switchings with mean switching currents of $I_{sw1}$ and $I_{sw2}$, respectively.
The shape of $P(I)$ clearly tended to be sharper with decreasing temperature~\cite{Matsu}.
This trend saturated around 0.5K, as clearly observed in the temperature dependence of $\sigma_{I}$ shown in Fig. 3. 
Figure 3 also shows a plot of $\sigma_{I}$ for the second switching. 
When the temperature was higher than the crossover temperature ($T^*_1\sim0.5K$ for the first switching and $T^*_2\sim6.5K$ for the second), both switchings followed a $T^{2/3}$ dependence, indicating a dominant TA process in the switching.
The saturation at $T^*_1$ in the first switching indicates the crossover of switching mechanism from the TA to the MQT ~\cite{Remark2}.
Actually, the measured $T^*_1$ is similar to previously reported values~\cite{Inomata,Jin}.
On the other hand, the significantly higher $T^*_2$ than $T^*_1$ suggests the presence of different origins for the second switching.
Since evaluated temperature dependence of the critical current for first switching $I_{c1}$ and for second switching $I_{c2}$ suggests the presence of interaction between the two junctions, we will discuss this topic in section 4.2.

\par
\subsection{First switching}
%
The dynamics of the first switching were analyzed here to determine the magnitude of the damping from experimental results shown in Fig. 3. 
Similar saturation behavior of $\sigma_{I}$ due to the MQT and relatively weak dissipation behavior in Bi2212 IJJs have already reported in ref. ~\citen{Inomata}.
Here we mainly focus on two points which have not been clarified in the previous work;
i) the damping effect evaluated from the switching properties above the crossover temperature,
ii) the influence of nodal QPs of the $d$-wave superconductivity on the switching properties from the switching properties below the crossover temperature with high precision measurements.
As a whole, we will present a consistent analysis on the switching dynamics of the IJJs
in terms of the underdamped JJ behaviors for the whole temperature range.
\par
%
%
Clarifying the damping effect is crucial for understanding the dynamics of a JJ. 
According to the Caldeira-Leggett formula ~\cite{CL}, the MQT rate is suppressed by the damping effect.
Thus the magnitude of the damping can be estimated from the crossover temperature and the saturation value of $\sigma_{I}$ ~\cite{Voss}.
However, such analysis is not adequate for the present case because the estimation of Q factor based on this method is sensitive to the capacitance, and the capacitance is roughly estimated from the size of the junction.
Another possible way to estimate the Q factor is to evaluate the gradient of $I$-$V$ characteristic inside the gap.
However this evaluation cannot be validated, because the experimental $I$-$V$ characteristics is influenced by the AC Josephson effect in the absence of applied field~\cite{Remark2,Silvestrini}.
Also the gradient of $I$-$V$ near zero-bias voltage is hard to be evaluated due to the retrapping effect. 
Moreover, the relation between the $I$-$V$ characteristics and the Q factor is unclear when the $I$-$V$ characteristics inside the gap show strong non-linearity due to the $d$-wave superconductivity, which is the condition in our study.
Therefore, our approach was to extract the Q factor from the TA process in which the escape rate depends on the damping, because the escape rate reflects the distribution of particles confined in the cosine potential.
Based on the analyses described in refs.~\citen{Devoret,Wallraff}, the escape temperature $T_{esc}$ of the TA for the first switching can be calculated by setting $a_t=1$ in Eq. 2 and assuming the cosine potential of Eq. 1~\cite{Kranov-TA}.
%
%
%
Figure 4 shows the $T_{bath}$ dependence of $T_{esc}$ obtained using the least squares method.
Below $T^*_1$, $T_{esc}$ was constant, reflecting the temperature-independent switching of the MQT.
Above $T^*_1$, although $T_{esc}$ was approximately $T_{bath}$, there was a non-negligible discrepancy between $T_{bath}$ and $T_{esc}$, and this discrepancy tended to increase with increasing $T_{bath}$.
This discrepancy was not due to measurement error or to heat generation at the contact electrodes.
Such discrepancy is similar to that observed in high-Q junctions of metal superconductors reported in ref.~\citen{Wallraff}.
The most plausible origin is the correction in the escape rate due to the deviation in $a_t$ from unity when the IJJ is in the underdamped regime.
%
The Q factor of the junction can be evaluated by assuming the following relation~\cite{Wallraff,Kramers,Hanggi},
\begin{eqnarray}
a_t=\frac{4}{(\sqrt{1+(Q k_b T/1.8 U_0)}+1)^2}.
\end{eqnarray}
Figure 4(b) shows the obtained Q factor in the temperature range from 0.6K to 5K.
Although the results were somewhat scattered due to statistical error, the Q factor was approximately $70\pm20$, indicating that the IJJ was clearly in the underdamped regime.
This Q factor is somewhat higher than that obtained by an YBa$_2$Cu$_3$O$_{7-\delta}$ junction evaluated using microwave-assisted tunneling, but comparable to that obtained by high quality metal superconductor JJs~\cite{Martinis1,Wallraff,Inomata2}.
\par
%
In the following, the dynamics of the first switching is shown to be consistent with that of a conventional underdamped JJ without the influence of the nodal QPs of the $d$-wave  for all temperature range below 5K.
First of all, the Q factor above the crossover temperature $T^*_1$ should be an exponentially decaying function of temperature if the Q factor is originated by the nodal QPs of the $d$-wave~\cite{HKash2}.
This is because the amount of QPs contributing to the damping increases with increasing temperature.
The expected Q factor in such situation is the order of $10^5$ at 0.5K according to ref.~\citen{HKash2}.
On the other hand, the experimentally detected Q factor shown in Fig. 4(b) is approximated by by Q=69.3+0.8$T$.
This value is high enough to be regarded as underdamped, but both the magnitude and the temperature dependence are quite different from those expected for the nodal QPs damping. 
Secondly, it is accepted that the crossover temperature is the function of damping as shown in ref.~\citen{CL}.
The experimental result on $T^*_1$ of about 0.5K is consistent with the underdamped theoretical value of 0.445K. 
Thirdly, although the Q factor below $T^*_1$ is not directly evaluated in the present experiments, $\sigma_{I}$=50-52nA are consistent with estimated value of 51.1nA ($I_c$=21.6$\mu$A, $C$=105fF, $\epsilon _{r}$=7.16) obtained from the underdamped MQT theory~\cite{CL}.
Also the lack of the temperature dependence of the $\sigma_{I}$ is inconsistent with $\sigma_{I}(T)- \sigma_{I}(0)\propto T^3$ which is the theoretical temperature dependence if the nodal QPs of the $d$-wave is the dominant origin of damping~\cite{Kawabata}.
This fact indicates that the nodal QPs are not the dominant origin of the damping in the MQT region.
The lack of the dominant influence of the nodal QPs of the $d$-wave on the MQT process is consistent with the theoretical estimation of ref.~\citen{Kawabata}.
As the whole, the dynamics of IJJ is consistent with that of an underdamped IJJ, and the dominant origin of the residual damping is not due to nodal QPs but to extrinsic factors such as microshorts in the IJJ or coupling to the environment via the signal lines.
Therefore, the Q factor can be improved by refining the junction fabrication process and by isolating the signal lines with proper microwave engineering similar to the technique used to improve the Q factor in metal superconductor junctions~\cite{Martinis2,Martinis3}.
\par
\subsection{Second switching}
%
In this section the properties of the second switching were studied by taking account of the self-heating effect and of the stack structure.
Although the junctions in the IJJs are capacitively coupled with each other~\cite{Koyama}, it is still unclear how this coupling affects the switching dynamics of each junction~\cite{Remark1}.
This is an important issue because weakly coupled JJs are promising candidates for multi-qubit applications~\cite{Martinis3,multiqubit}.
We analyze the switching current distribution under the influence of dissipative current based on the conventional TA model ~\cite{Kramers,Hanggi}.
The advantages of the present method are simple device structure and high sensitivity of the switching current distribution on the temperature.
\par
Three notable features were observed in the result of second switching and details are discussed below:\\
1. The detected crossover temperature ($T^*_2$) was 6.5K (Fig. 5) which is far higher than theoretically expected value of 0.74K.\\
2. The reversal of the relative magnitude of the mean switching current was detected, that is, $I_{sw1} >I_{sw2}$ below 1.1K and $I_{sw1} < I_{sw2}$ above 1.1K (Fig. 6).
The relation of $I_{sw1} >I_{sw2}$ is inconsistent with the naive expectation that each junction in the stack switches to the finite voltage state on the order of smaller $I_{c}$ when the bias current is gradually increased. \\
3. Although it is accepted that the temperature dependence of $I_{c}$ follow AB formula in the cases of IJJs, both of the temperature dependences of $I_{c1}$ and $I_{c2}$ in Fig. 6 are inconsistent with Ambegaokar-Baratoff (AB) formula. \\

%
%
\par
%
For the barrier potential of the second switching, we assumed a simple tilted cosine potential as described in Eq. 1.
Figure 5 shows the temperature dependence of $T_{esc2}$ for the second switching obtained using the least mean squares method.
The $T_{esc2}$ shows saturation behavior at about $T^*_2\sim6.5K$.
Because the calculated $T^*_2$ from the TA to the MQT for the second switching is 0.74K (=5$\hbar\sqrt{\frac{2\pi I_c}{\Phi_0 C}}$/36$k_B$ ), the observed $T^*_2$ is too high to expect the transition to the quantum regime.
Furthermore, even above $T^*_2$, significant discrepancy is evident between $T_{bath}$ and $T_{esc2}$.
Such large discrepancy is inconsistent with conventional JJ behavior, even assuming underdamped conventional JJs.
However, these anomalies are understood in terms of the self-heating effect, which is due to the poor thermal conductivity of Bi2212~\cite{sh2,sh3,sh4,sh5,insitu}.
When all the junctions are in the zero-voltage state, no dissipative current exists in the IJJ, and thus the effective electron temperature in the stack should be equal to the $T_{bath}$.
Whereas the influence of the dissipative current cannot be neglected when the first junction has been transited to the finite voltage states.
The most plausible explanation is that the measured $T_{esc2}$ directly corresponds to the effective electron temperature in the stack and the switching events are induced by the enhanced thermal fluctuation (referred to as self-heating explanation)~\cite{aa}.
Along this explanation, the evaluation of switching distribution can be regarded as an $in$-$situ$ measurement of the effective electron temperature in the stack~\cite{sh5,insitu}.
Actually, similar discrepancy between $T_{esc2}$ and $T_{bath}$ has been reported for the stacked structure as well as mesa structures in which QPs were injected into the JJ from a metal electrode placed near the junction~\cite{Kranov-TA}.
The uniqueness of the present work is that the sensor used to detect the effective temperature (secondly switched junction) locates quite close to the heating spot (firstly switched junction), and then the influences of nonequilibrium effect cannot be neglected in addition to the diffusive joule heating.
\par
%
In order to compare the present results with those in previous works, we evaluate the thermal resistance $R_{th} \equiv \Delta T/P$ 
where $\Delta T$ and $P$ are the change of temperature and the injected power, respectively.
The value of $R_{th}$ in our measurement is given by 2.3 $\times 10^{7}$(K/W) ($\Delta T \sim $8.5K, and $P \sim $ 17 mV$\times$ 21.6$\mu$A), which is far higher than those detected in mesa ~\cite{insitu}.
Also the $\Delta T$ estimated based on the equation presented in ref.~\citen{sh4} taking account of mesa structure gives 0.36K by assuming $\kappa_{ab}$ at $<$1K is 1W/Km and $\kappa_{ab}/\kappa_{c} \sim $10 ~\cite{Ando}, which is far smaller than the present results.
These serious discrepancies reflect the high sensitivity of the stacked junction structure for the self-heating as compared to that of mesa.
Whereas in the case of the stacked structure similar to that used in the present work~\cite{sh5}, the relatively high thermal resistance of  $\sim$2.8$\times 10^{5}$(K/W) with $\Delta T \sim$31K (110$\mu$W) was obtained.
Even taking account of the facts that $\kappa_c$ at 35K is about 3 times larger than $\kappa_c$ at 8.5K,
and that the distance between the sensor and the heat spot in ref.~\citen{sh5} is estimated more than ten times as far as that in our case, the evaluated thermal resistance in our experiments is effectively several times larger than that obtained in ref.~\citen{sh5}.
However, such difference is not  unreasonable because the device used in ref.~\citen{sh5} implemented two side bars and they worked as additional heat sinking paths.
Another important factor for the relatively enhanced self-heating effect in our measurement may stem from the position of the "temperature sensor". 
In the present measurement, the temperature was monitored by the secondly switched junction that locates quite close (1.5-9nm) to the firstly switched junction. 
Therefore, in addition to the influence of the simple joule heating, serious influences by the ballistic phonon and by the nonequilibrium quasiparticle distribution (NQD) due to quasiparticle injection at firstly switched junction cannot be neglected.
This is because the distance of those two junctions was less than the c-axis quasiparticle relaxation length (about 16nm estimated in ref.~\citen{You}) as well as the mean free path of phonons.
Although the influence of the NQD on the switching current distribution is not clear at present, it can be approximated as elevated electron temperature similarly to the usual treatment of nonequilibrium superconductivity ~\cite{Parker}. 
Therefore, we ascribe that the origin of the enhanced $\Delta$T in the second switching to the self-heating effects, that is, the sum of the simple Joule heating, the ballistic phonon and NQD~\cite{higher}.
\par
%
%
The TA process analysis treats the mean switching current ($I_{sw1}$, $I_{sw2}$) and the critical current ($I_{c1}$, $I_{c2}$) as independent parameters.
Figure 6 summarizes the analysis results for the IJJs studied here. 
At the first switching, the difference between $I_{c1}$ and $I_{sw1}$ was less than 1$\mu$A below $T^*_1$, and the difference rapidly increased when $T_{bath}$ was increased from $T^*_1$.
In contrast, at the second switching, the difference between $I_{c2}$ and $I_{sw2}$ was almost at 6$\mu$A for all $T_{bath}$ below 5K because of the increased effective electron temperature due to the heating effect.
The difference in the temperature dependence between $I_{sw1}$ and $I_{sw2}$ causes the reversal of the relative magnitudes of currents between the first and second switchings at about 1.1K (=$T_{CR}$).
%
The fact that $I_{sw1} > I_{sw2}$ has also been reported in ref.10. 
\par
%
Here we discuss the anomaly of the temperature dependences of $I_{c1}$ and $I_{c2}$ in Fig. 6.
It is accepted that the temperature dependence of $I_{c}$ follows AB formula as suggested by Krasnov $et$ $al$~\cite{Kranov-TA}. 
Based on the AB formula, the critical current keeps constant value ($0.9999<I_{c}(T)/I_{c}(T=0)<1.0$) in the temperature range $0<T<T_{c}/5$, therefore $I_{c}$ should be constant in the present experimental temperature range shown in Fig. 6.
However, $I_{c2}$ in Fig. 6 is gradually decreasing as the temperature is increased. The amount of change is $I_{c2}$(T=5K)/$I_{c2}$(T=60mK)=0.969, which is far larger than that expected from AB formula. 
Furthermore, $I_{c1}$ in Fig. 6 is increasing as the temperature is increased with the amount of change of $I_{c1}$(T=5K)/$I_{c1}$(T=60mK)=1.016. 
The fact that the $I_{c}$ is an increasing function of temperature is anomalous because it is inconsistent with conventional theories for weak-link Josephson junction~\cite{Barone} as well as AB formula.
One possible explanation for these anomalies is that $I_{c}$ of the IJJ in the stack is modified from that of isolated IJJ due to the coupling effect~\cite{Koyama}, thus it could have different temperature dependence from that of isolated IJJ~\cite{Remark1}.
However we don't have detailed model to explain the mechanism of the anomalous temperature dependence of $I_c$ in the stacked junction at present.
\par
%
Apart from above-mentioned anomalies, the dynamics of the second switching is rather consistent with the conventional TA model of an isolated JJ.
As shown in Fig. 7, $P(I)$ in the second switching consistently follows the theoretical curves of the TA model, and no additional structures, such as peaks and dips, were found within the present resolution.
This consistency suggests that the oscillation of the macroscopic phase (AC Josephson effect) at the firstly switched junction does not seriously modifies the second switching properties\cite{peak}.
On the other hand, the influence of the self-heating effect is serious for multiqubit application of the junction stack because the self-heating might destroy quantum coherence of the others.
It is important to check whether this problem is peculiar to the present stacked structure or not.
The Joule heating may be reduced in the case of mesa structure because of the presence of the heat sinking path at the topmost Au layer.
Also the miniaturization of the junction size would be effective in order to reduce the influence of the Joule heating.
If we adopt the similar size dependence for the present junction to that of mesa~\cite{sh4}, the junction size of 47nm gives the heat generation of 0.5K by assuming that the current density is size independent.
However, it is not clear at present whether the self-heating can be reduced by the miniaturization of the junction size 
or the implementation of the additional heat-sinking path, because the non-equilibrium effect cannot be avoided by these modifications.
Further studies are needed in order to realize the multiqubit  in a single IJJs.
\par
%
%
\section{Summary}
%
In this study, high quality Bi$_{2}$Sr$_{2}$CaCu$_{2}$O$_{8+\delta}$ intrinsic Josephson junctions (IJJs) with large hysteresis in the $I$-$V$ characteristics were fabricated.
The escape rate measurements for the IJJs clarified the switching dynamics of the IJJs as follows.
At the first switching, both the escape temperature and the width of the switching probability distribution saturated below 0.5K, indicating the transition of the switching mechanism from the thermal activation (TA) to the macroscopic quantum tunneling (MQT).
The high quality (Q) factor (70$\pm$20) obtained for the first switching reveals underdamped features, despite possible damping due to the nodal quasiparticles of the $d$-wave and the stack structure. 
Because the dominant origin of the damping is probably not due to the inherent properties of the IJJs, higher Q factor might be achieved by improving the connection of the signal lines by using proper microwave engineering.
At the second switching, however, the escape temperature is almost temperature independent below 6.5K, indicating significant influence of the self-heating effect.
The switching of the first junction into a finite voltage state increased the effective electron temperature, and the escape temperature directly corresponds to the effective electron temperature in the stack.
Although the self-heating explanation reasonably describes the enhanced fluctuation in second switching, the anomalous behaviors in the temperature dependence of critical current remain unsolved.
The heating effect might destroy the quantum coherence in multiqubit of IJJs, however, this problem will be overcome if the IJJ with junction size of several tens of nanometers would be fabricated.
In conclusion, the results from this study strongly indicate that high-$T_c$ superconductor junctions can be used in quantum information device applications.

\par
\section*{Acknowledgements}
We thank K. Inomata, K. Oka, T. Kato, K. Semba, and H. Yamaguchi for fruitful discussions.

This work is supported by Grant-in-Aid for Scientific Research in the Priority Area
"Novel Quantum Phenomena Specific to Anisotropic Superconductivity"
(Grant No. 17071007) from the Ministry of Education, Culture, Sports,
Science and Technology of Japan. 
%
%
%

%
\newpage
\begin{figure}[hb]
\begin{center}
\includegraphics[width=0.8\linewidth]{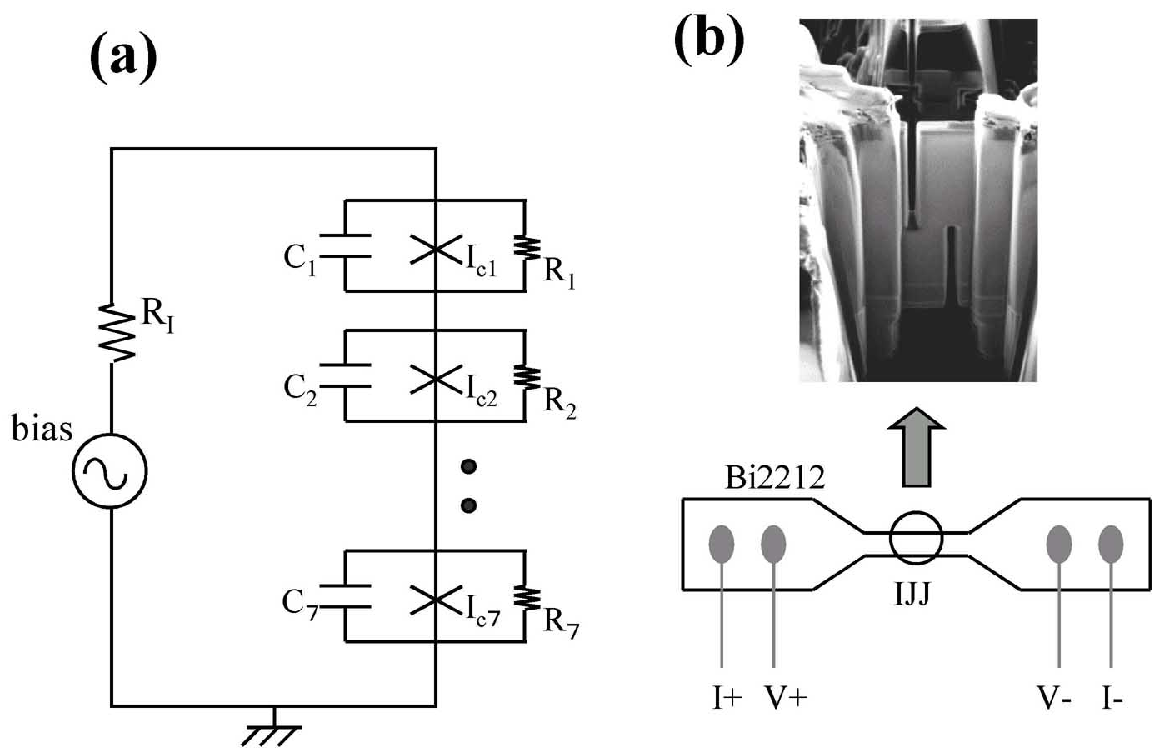}
\end{center}
\caption{
(a) Equivalent circuit model of the stacked IJJ structure.
The stacked IJJ includes seven JJs in a stack, and the RCSJ model represents each junction.
(b) Schematic and scanning ion micrograph image of the stacked IJJ structure. 
}

\end{figure}

\begin{figure}[htb]
\begin{center}
\includegraphics[width=0.8\linewidth]{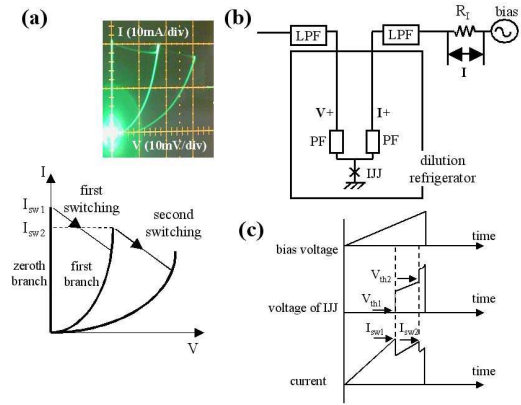}
\end{center}
\caption{ 
(color online). 
(a) Schematic and photograph of the $I$-$V$ characteristics of the stacked IJJ at 55mK.
Large hysteresis confirms the high quality of the IJJ.
The first and second switchings shown here respectively correspond to the transition from the zero to the first branch and from the first to second branch.
$I_{sw1}$ and $I_{sw2}$ indicate the switching current for the first and second switchings, respectively.
The relation $I_{sw1}>I_{sw2}$ is clearly evident in the I-V characteristics (see text).
(b) Schematic of the electronics used to measure the first and second switchings.
Bias current was introduced from the bias circuit to the junction through a bias resister ($R_{I}$) and filters.
Both the current bias line ($I_+$) and voltage monitor line ($V_+$) were carefully filtered by a low-pass filter (LPF) and powder filter (PF).
Other connections ($I_-$, $V_-$) were connected to the sample holder.
(c) Schematic of time-oscillograms used in the present measurements.
}
\label{f2}
\end{figure}

\begin{figure}[htb]
\begin{center}
\includegraphics[width=0.8\linewidth]{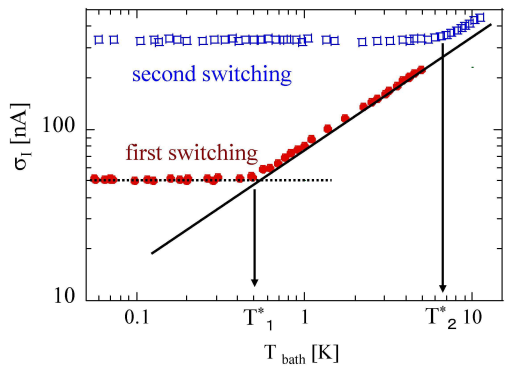}
\end{center}
\caption{
(color online). 
Temperature dependence of $\sigma_{I}$ defined by the standard deviation of the switching probability distribution P(I) for the first switching (red symbols) and second switching (blue symbols).
Solid line represents a guideline for $T^{2/3}$ dependence, and dotted line represents the theoretical value (51.1nA) for the MQT without damping.
}
\label{f3}
\end{figure}
%
\begin{figure}[htb]
\begin{center}
\includegraphics[width=0.8\linewidth]{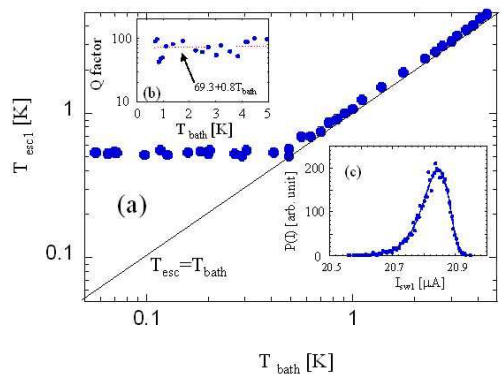}
\end{center}
\caption{
(color online). 
Bath temperature dependence of the escape temperature for the first switching ($T_{esc1}$) based on the TA model with $a_t$=1.
Solid line corresponds to $T_{esc}=T_{bath}$.
Inset (b) shows bath temperature dependence of the Q factor for the first switching based on the TA model with correction of the thermal prefactor taken into account (see text).
Dotted line represents fitting to the distribution (Q=69.3+0.8$T_{bath}$, indicating that the Q factor was relatively independent of bath temperature.
Inset (c) shows switching probability distribution $P(I)$ at $T_{bath}=70mK$ and the theoretical fitting ($I_c$=21.637$\mu$A, C=105fF, Q=70) based on Eq. (3).
}
\label{f4}
\end{figure}
%
%
\begin{figure}[htb]
\begin{center}
\includegraphics[width=0.8\linewidth]{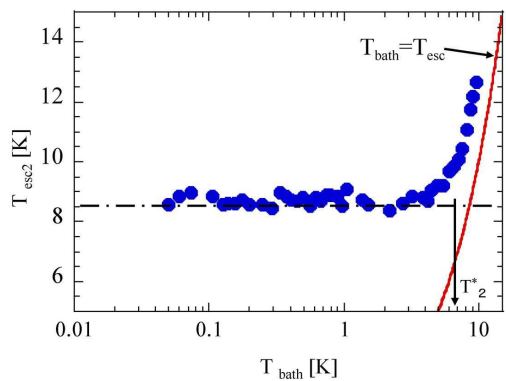}
\end{center}
\caption{
(color online). 
Bath temperature dependence of the escape temperature for the second switching ($T_{esc2}$).
Solid line corresponds to $T_{esc2}=T_{bath}$, and the dashed line corresponds to the saturation temperature ($T_{esc2}\simeq$ 8.5K) of the escape temperature at $T_{bath} < T^*_2$.
}
\label{f5}
\end{figure}
%
\begin{figure}[htb]
\begin{center}
\includegraphics[width=0.8\linewidth]{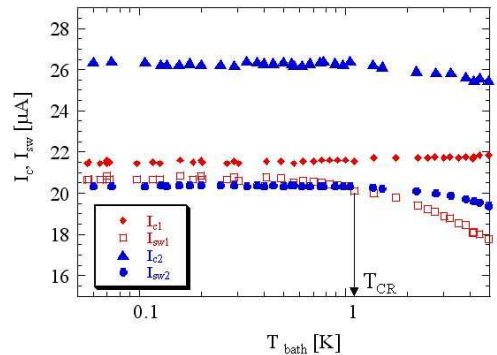}
\end{center}
\caption{
(color online). 
Bath temperature dependence of the mean switching current $I_{sw1}$ and $I_{sw2}$ extracted from the experimental data and the critical current $I_{c1}$ and $I_{c2}$ from the fit to the TA model for the first switching (red symbols) and second switching (blue symbols).
The crossover of the magnitude of $I_{c1}$ and $I_{c2}$ occurs at $T_{CR}$ ($\simeq$ 1.1K ).
}
\label{f6}
\end{figure}
%
%
\begin{figure}[htb]
\begin{center}
\includegraphics[width=0.8\linewidth]{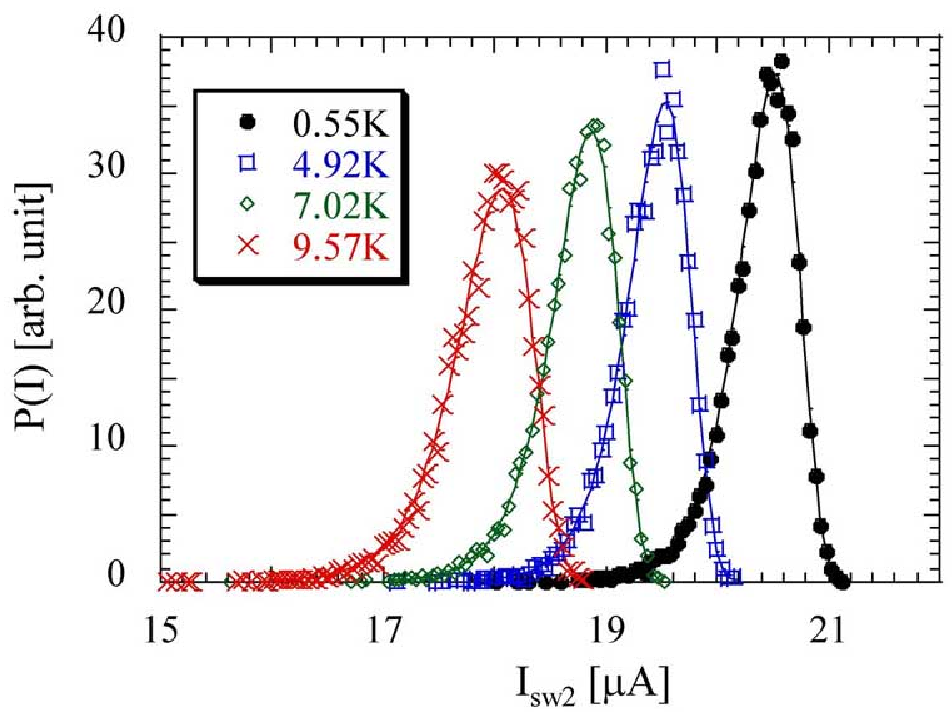}
\end{center}
\caption{
(color online). 
Switching probability distribution $P(I)$  for the second switching in the temperature range between 0.55K and 9.57K.
Symbols represent experimental data, and solid lines represent the theoretical fitting based on Eq.(2).
}
\label{f7}
\end{figure}


\begin{thebibliography}{99}
\bibitem{Remark1} 
When the finite coupling exists between the junctions in the stack, the dynamics of each junction is modified from that of an isolated junction.
The junction parameters used in the present analysis should not be those for an isolated JJs but for modified Josephson junctions with the effect of coupling taken into account.
As described in the text, an analogy is found in a weakly coupled two-coordinate oscillator model.
When we assume two identical oscillators, the normal modes of the coupled oscillators exhibit so-called "bonding antibonding splitting".
In such situation, the variation of coupling interaction strength yields contradicting dependences on the frequency of the two modes, that is; the frequency of a normal mode increases, while the frequency of the other normal mode decreases with increasing the coupling interaction strength.
This model may intuitively explain the increased $I_{c1}$ with decreased $I_{c2}$ when the temperature is increased as shown in Fig. 6.
However, this explanation is oversimplified because the present stack includes seven junctions thus the coupling of the modes exhibits more complex behavior.
Also the origin of temperature dependence of the coupling interaction strength is unclear at present.
\bibitem{Remark2} 
Because the present junction was unstable in the thermal cycling even up to 100K (just above $T_c$), the same junction condition could not be reproduced once the trapping occurred.
Thus, magnetic field modulation experiments were not performed on the present junction.
We did confirm, however, that the noise amplitude (less than 10nA in 1$\sim$10MHz) evaluated for the present electric circuit was small enough to detect the saturated $\sigma(I) \sim $ 51nA.
\bibitem{Inomata2}  The Q factor $\sim10^2$ has been suggested for Bi2212 estimated by the gradient of the I-V characteristics in ref.~\citen{Inomata}.
\bibitem{aa} The estimated heating at the end of seventh branch for our case (bias condition of 119 (17$\times$7) mV and 36$\mu$A, and $\kappa$ at 35K is assumed as about 3 times larger than $\kappa$ at 8.5K) is around 35K. 
This level of heating is never inconsistent with the observation of good periodicity of branches, because Ambegaokar-Baratoff formula suggests that the critical current is largely suppressed only when $T>2/3T_c$ (60K in the present case).
\bibitem{higher} We didn't evaluate the higher order switching properties in the present junction based on technical problems, such as, the limitation of electronics (saturation of preamplifier), the stability of the temperature.
%
Also the order of switching was tend to be unstable for higher bias voltage, that is, the switching of A-B-C-D mixed with A-B-D-C (A, B, C, D are junction in the stack, respectively). 
%
The complex switching order interrupt the evaluation of correct switching probability distribution for higher order switchings.
\bibitem{peak}
After the first switching, the firstly switched junction irradiates photons due to the AC Josephson effect.
If the second switching is influenced by these photon, the addtional peak formation or peak shift due to the photon assisted tunneling effect  would be expected in the switching probability distribution such as those reported in refs. \citen{Bauch} and \citen{Jin}.
\end{thebibliography}
\end{document}